\documentclass[twocolumn,trackchanges]{aastex631}
\newcommand{\bprp}{$G_{\rm BP}-G_{\rm RP}$}
\newcommand{\rvar}{$R_{\rm var}$}
\newcommand{\logprot}{log$_{10}P_{\rm rot}$}
\newcommand{\prot}{$P_{\rm rot}$}
\newcommand{\NZTFoutlier}{529}
\newcommand{\NASASSNoutlier}{17}
\newcommand{\NZTFoutlierR}{150}
\newcommand{\NASASSNoutlierR}{9}
\newcommand{\NZTFvet}{18,614}
\newcommand{\NASASSNvet}{1,018}
\newcommand{\NClusters}{1,368}

\begin{document}

\title{20,000 Rotation Periods in Nearby Open Clusters and Moving Groups from ZTF and ASAS-SN}

\newcommand{\osu}{Department of Astronomy, The Ohio State University, Columbus, 140 W 18th Ave, OH 43210, USA}
\newcommand{\ccapp}{Center for Cosmology and Astroparticle Physics (CCAPP), The Ohio State University, 191 W. Woodruff Ave., Columbus, OH 43210, USA}

\author[0000-0003-4769-3273]{Yuxi(Lucy) Lu}
\affiliation{\osu}
\affiliation{\ccapp}



\begin{abstract}
Open clusters provide uniquely powerful laboratories for studying stellar evolution because their members share a common age and chemical composition. 
Stellar rotation of low-mass GKM dwarfs in open clusters provides direct empirical constraints on magnetic braking, with implications for stellar interiors, magnetic activity, and star--planet interactions. 
However, existing cluster rotation samples are dominated by space-based observations, whose limited spatial coverage and short observing baselines can make it difficult to construct an unbiased sample.
Long-baseline ground-based time-domain surveys therefore provide an important complement to space-based photometry.
I present a catalog of $\sim$20,000 periodic photometric signals, including stellar rotations, eclipsing-binaries, and classical pulsators, measured from the Zwicky Transient Facility (ZTF) and the All-Sky Automated Survey for Supernovae (ASAS-SN) for members of nearby open clusters and moving groups within 1.5 kpc. 
This catalog provides a new empirical foundation for studying the mass, chemical composition, and age dependence of stellar angular-momentum evolution and for testing models of magnetic braking across a broad range of stellar masses and evolutionary states.
\end{abstract}



\section{Introduction} \label{sec:intro}
Open clusters are co-evolved stellar populations that form from the same molecular cloud, sharing a common age and chemical composition and providing valuable laboratories for studying stellar evolution across different masses and evolutionary stages.
Because cluster members share approximately the same age, chemical composition, and birth environment, open clusters provide powerful empirical benchmarks for star formation \citep[e.g.,][]{Phelps1997}, stellar age calibration \citep[e.g.,][]{Soderblom2010, Jeffries2023, Bouma2023, Lu2024, vanlane2025, Lu2026}, stellar evolution \citep[e.g.,][]{Mermilliod1981, VandenBerg2004}, atomic diffusion \citep[e.g.,][]{BertelliMotta2018, Souto2019, Souto2026}, binary evolution \citep[e.g.,][]{Hurley2005, Mathieu2009}, and chemical tagging \citep[e..g,][]{Freeman2002, Bovy2016, Ness2018}. 

Rotation periods of open-cluster members are particularly important for empirically understanding the evolution of stellar angular momentum and magnetic activity, as well as for constraining theoretical models of magnetized stellar winds \citep[e.g.,][]{Skumanich1972, Kawaler1988, Barnes2003, Matt2015, Garraffo2018, Spada2020, Cao2022, Spada2026}. 
This evolution is closely connected to the structure and evolution of stellar interiors, including the transition between different dynamo regimes and the coupling of stellar rotation to magnetic activity \citep[e.g.,][Lu et al. in prep.]{Noyes1984, Donati2009, Charbonneau2010}. 
Stellar rotation also has important consequences for star–planet interactions since the magnetic and high-energy environments of host stars evolve as stars spin down, influencing planetary atmospheric escape, habitability, and the long-term evolution of planetary systems \citep[e.g.,][]{Khodachenko2007, Jackson2012, Airapetian2016, See2025}.

Measurements of rotation periods across a wide range of ages have revealed that stellar spin-down is not described by a single universal relation, with departures from simple gyrochronology emerging for both young stars and old field populations \citep[e.g.,][Lu et al. in prep.]{Curtis2020}. 
In particular, the discovery of weakened magnetic braking among older stars \citep[e.g.,][]{vanSaders2016}, stalled spin-down for K-dwarfs \citep[e.g.,][]{Curtis2020, Lu2022}, and unexpected rotational behavior of fully convective stars \citep{Lu2024_fc, Chiti2024, Newton2016}, have raised fundamental questions about how and when stars cease to follow the classical magnetic-braking relations. 
In addition, theoretical models and observations of field stars \citep[e.g.,][]{Amard2020, See2024} indicate that metallicity can affect a star's internal structure and, consequently, its angular momentum loss.
However, this effect has not been confirmed in clusters, where age and selection effects can be controlled for, because most existing cluster rotation samples are limited to nearby, approximately solar-metallicity clusters.
As a result, extending precise rotation-period measurements to a larger and more diverse sample of open clusters, particularly across a wider range of ages and metallicities, can provide critical constraints on the mass-, age-, and chemical-composition dependence of angular-momentum loss.

Current measurements of rotation periods in open clusters have been driven largely by space-based photometric surveys. 
The original Kepler \citep{kepler} mission provided an exceptional 4-year photometric baseline and high-precision, nearly continuous light curves. 
Several open clusters fall within the Kepler field, including NGC 6866, NGC 6811, NGC 6819, and NGC 6791, but rotation-period studies have been most successful for the relatively nearby NGC 6811. 
At a distance of $\sim$1 kpc, NGC 6811 is sufficiently close for Kepler to detect rotational modulation in a substantial population of its lower-mass members, enabling precise measurements of rotation periods and the first well-defined intermediate-age cluster rotation sequence \citep{Meibom2011}. 
The other Kepler-field clusters are substantially more distant, making their low-mass members considerably fainter and limiting the utility of Kepler photometry for constructing comparable rotation samples. 
Thus, despite its exceptional temporal baseline, the original Kepler mission provided detailed cluster rotation measurements for only a limited number of nearby systems.

Subsequent space missions have greatly expanded the number of clusters accessible to rotation studies. 
K2 \citep{K2} observed numerous nearby open clusters, including the Hyades and Praesepe  \citep[e.g.,][]{Rebull2017, Douglas2019}, but individual campaigns typically lasted only $\sim$70--80 days.
TESS  \citep{tess} provides nearly all-sky coverage and has enabled rotation measurements for many more clusters \citep[e.g.,][]{Long2023, Boyle2026}, but its nominal observing baseline is only $\sim$27 days per sector. 
These relatively short baselines become increasingly restrictive for older clusters, where slower-rotating stars are more common and are particularly important for tracing the long-term evolution of stellar angular momentum.
Even in younger clusters, where most stars are relatively rapid rotators, short observational baselines can systematically miss the population of slowly rotating stars (Lu et al., in prep.). 

Long-baseline, ground-based time-domain surveys such as Zwicky Transient Facility \citep[ZTF;][]{ztfdata, ztftime} and All-Sky Automated Survey for Supernovae \citep[ASAS-SN;][]{asassn2014} therefore provide an important complement to space-based photometry. 
In this paper, I present a catalog of $\sim$20,000 periodic photometric signals ($\sim$10,000 stars in the GOLD sample), including stellar rotation and eclipsing-binary variability, measured from ZTF and ASAS-SN for members of nearby open clusters and moving groups within 1.5 kpc. 
I use the membership catalog of \citet{Hunt2023}, in \autoref{sec:data_method}, I describe the data and methodology used to identify and measure these periodic signals. 
In \autoref{sec:results}, I present the resulting period and photometric-variability distributions and examine the properties of the detected populations.
In \autoref{sec:Uses}, we highlight several applications and possible future directions.

\section{Data \& Method} \label{sec:data_method}
\subsection{Cluster Membership \& Light Curve Data}
I adopt the cluster membership catalog of \citet{Hunt2023}, who identified cluster members in Gaia DR3 by applying the HDBSCAN clustering algorithm to 5D astrometric data, including parallax, proper motion, and Galactic longitude and latitude. 
I include all candidate members with membership probabilities $>0.2$ to prioritize completeness in my sample.

The irregular temporal sampling of ground-based photometric surveys can introduce aliases and systematic signals that may be mistaken for stellar variability. 
In particular, signals associated with short-period systematics can appear at longer apparent periods in the presence of uneven sampling. 
The more uniform temporal sampling provided by space-based observations offers an independent way to identify these ground-based failure modes (see \autoref{subsec:period_measure}).
Because many of the dominant systematic periods occur at $<10$ days \citep{Lu2022}, I use the TESS all-sky rotation-period catalog of \citet{Boyle2026} to identify, vet, and correct potential systematic periods in the ground-based measurements. 
I additionally compare my measurements against the compilation of space-based cluster rotation periods from \citet{vanlane2025}, which combines measurements from Kepler, K2, and TESS observations reported in the literature \citep{Meibom2015, Douglas2016, Barnes2016, Douglas2019, Douglas2024, Rebull2016, Rebull2017, Rebull2018, Curtis2019, Rampalli2021, GodoyRivera2021, Long2023}.
In addition, I included the moving group period measurements from \cite{Kounkel2022} that overlap with the \cite{Hunt2023} catalog.

\subsection{Rotation Period Measurements}\label{subsec:period_measure}
Because ZTF has sky coverage limited to declinations above $-25^\circ$, I use ZTF light curves whenever available. 
For cluster members at declinations below $-25^\circ$, where ZTF coverage is unavailable, I instead search for rotational variability using ASAS-SN observations.
For the ZTF sample, I restricted the search to stars with Gaia $G < 18$ mag, while the corresponding magnitude limit for ASAS-SN is $G < 16$ mag.

For ZTF, I adopt the Zubercal light curves \citep[ZTF Ubercal;][]{drake2025zubercal}\footnote{Available at \url{http://atua.caltech.edu/ZTF/Zubercal.html}.}, a recently recalibrated version of the ZTF photometry designed to improve the accuracy and stability of the standard PSF measurements. 
Zubercal performs a joint photometric calibration for all PSF measurements acquired in the same detector quadrant (1/64 of the full ZTF field of view), filter, and observing night. 
This procedure mitigates the substantial systematic residuals (reach up to 5\%) present in the standard IPAC-calibrated ZTF photometry available through IRSA.
I restrict my analysis to the $r$ band, which provides the largest number of epochs among the available ZTF filters. 
I search for periodic signals using the Lomb--Scargle periodogram only for stars with at least 50 measurements that do not exhibit photometric quality flags and have magnitude uncertainties below 0.1 mag.

For ASAS-SN, I obtain light curves through the ASAS-SN Sky Patrol \citep{Hart2023}. 
I only analyzed data in the $g$ filter as it has the largest number of available observations.
I restrict my analysis to stars with at least 50 good-quality photometric measurements and with magnitude uncertainties below 0.1 mag.

For each light curve, I search for periodic signals with periods between 0.2 and 100 days using the \texttt{astropy.LombScargle} implementation \citep{astropy:2013, astropy:2018, astropy2022}, with MJD, magnitude, and magnitude uncertainty as inputs. 
For the highest-power period identified by the periodogram, I calculate its false-alarm probability (FAP) using the corresponding \texttt{astropy} functionality. 
I then apply a series of quality cuts to select reliable periodic signals, requiring a maximum Lomb--Scargle power $>0.1$, a ratio of the maximum to median periodogram power $>10$ as a measure of signal significance, a light-curve baseline at least 5 times the detected period, and FAP $<0.1$.

For the remaining periodic signals, I identify potential failure modes of the Lomb--Scargle periodogram arising from uneven temporal sampling. 
Such sampling can introduce aliasing and spectral leakage, producing spurious peaks or causing the periodogram to select an incorrect period instead of the true rotation period \citep{VanderPlas2018}.
I characterize these aliases using
\begin{equation}\label{eq:1}
P_{\rm obs} = \left| \frac{m}{P_{\rm true}} + \frac{n}{\delta P} \right|^{-1},
\end{equation}
where $P_{\rm obs}$ is the observed alias period, $P_{\rm true}$ is the true period, $m$ and $n$ are integers, and $\delta P$ represents a characteristic sampling period or the period of a systematic signal.

I determine the characteristic sampling periods from the spectral window function over periods between 0.2 and 1000 days. 
The spectral window is constructed by assigning unit flux to each observation and computing the Lomb--Scargle periodogram of the resulting time series. 
I identify peaks exceeding $3\sigma$ above the median periodogram power using 1000 randomly selected light curves from ZTF and ASAS-SN separately. 
This procedure identifies characteristic sampling periods of 0.33, 0.50, 1.00, 363.67, and 444.48 days for ZTF, and 0.25, 0.33, 0.5, 1, and 363.67 days for ASAS-SN. 
I additionally include known sources of periodic systematics, such as the $\sim$29.5-day lunar cycle, based on previous studies and inspection of the ensemble period distribution.
I use these characteristic periods to identify and remove potential aliases. 
Specifically, I exclude periods within 10\% of the known systematic periods. 
For signals that fall along predicted alias relations involving a characteristic sampling period (identified as the log difference between the frequencies), I compare the detected period with available literature measurements and correct the period when the literature value provides evidence for the true rotation period.

\begin{figure*}
    \centering
    \includegraphics[width=\linewidth]{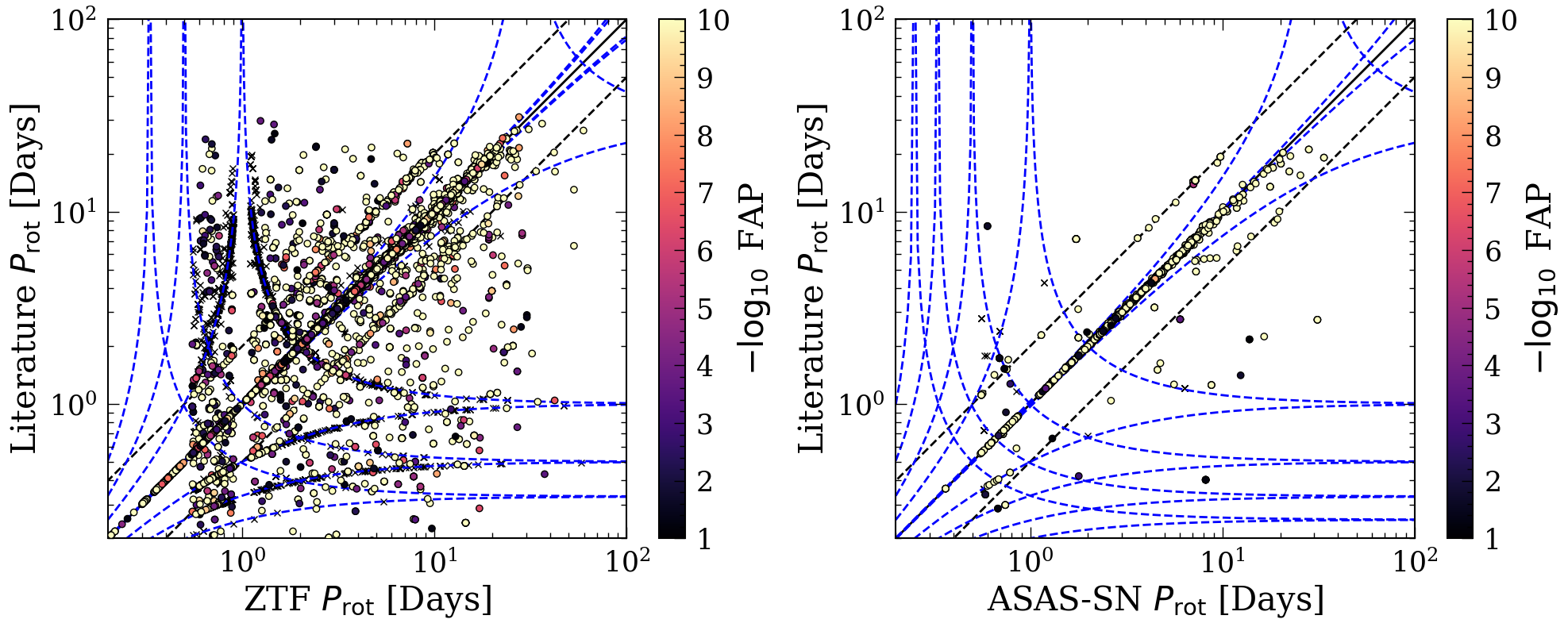}
    \caption{Comparison between rotation periods measured from ZTF and ASAS-SN in this work and periods reported in the literature. 
    The solid black line indicates the one-to-one relation, and the dashed black lines show the half-period harmonics. 
    Blue dashed lines indicate the alias relations predicted by \autoref{eq:1} based on the characteristic sampling periods. 
    Crosses mark sources identified as lying along these failure modes, for which I adopt the literature period. 
    Points show the accepted periods, colored by $-\log_{10}(\mathrm{FAP})$, with larger values indicating more statistically significant periodogram peaks.
    After vetting and correcting the detected periods, 94\% of the ZTF and 98\% of the ASAS-SN measurements are consistent with the literature periods or their half- or double-period harmonics.}
    \label{fig:1}
\end{figure*}

\autoref{fig:1} compares the periods measured from ZTF and ASAS-SN in this work with rotation periods reported in the literature. 
The solid black line indicates the one-to-one relation, while the dashed black lines mark the half-period harmonics.
The blue dashed lines show the alias relations predicted by \autoref{eq:1}. 
Crosses denote sources whose measured periods are identified as lying along one of these alias relations; for these sources, I adopt the literature period as the correct rotation period. 
The remaining points show the accepted periods from my analysis, with the color scale representing $-\log_{10}(\mathrm{FAP})$, where a larger number represents a smaller FAP value.
Larger values correspond to lower false-alarm probabilities and therefore more statistically significant periodogram peaks.
After correcting for the identified failure modes, which affect 13\% and 1.6\% of the ZTF and ASAS-SN measurements, respectively, 94\% of the ZTF and 98\% of the ASAS-SN measurements are consistent with the literature periods or their half- or double-period harmonics.

Because the sample is too large to visually inspect every detected rotation period, I manually vet $\sim1000$ period outliers (as defined below) as well as stars in clusters older than 1 Gyr. 
I prioritize these sources because period outliers are more likely to have incorrect periods, while older clusters are relatively rare and their slowly rotating stars are more difficult to detect and measure.
I identify outliers using a Gaussian Mixture Model (GMM) constructed separately for the ZTF and ASAS-SN samples, with \bprp, $M_G$, \logprot, $\log_{10}(\mathrm{Age})$, and $\log_{10}(R_{\rm var})$ as input features. 
For each survey, I select sources in the lowest 10\% of the GMM probability distribution among stars with $10^{-4}<\mathrm{FAP}<0.1$. 
I visually inspect the resulting \NZTFoutlier\ ZTF and \NASASSNoutlier\ ASAS-SN candidates and retain \NZTFoutlierR\ and \NASASSNoutlierR\ sources, respectively, with clear and reliable rotation-period measurements based on their folded light curves.
These results are stored in the \texttt{Inspection} column in the final rotation catalog.
Finally, for Praesepe and the Hyades, I adopted the rotation-period measurements from Lu et al. (in prep.), for which I visually inspected all stars. 
These measurements are in good agreement with those from the independently vetted sample from this work.
Only stars in these two clusters with a field contamination probability $<$0.01 from Lu et al. (in prep.) are included.

The final catalog contains \NZTFvet\ periodic signals from ZTF and \NASASSNvet\ from ASAS-SN. 
The column descriptions are shown in \autoref{tab:1}.
These measurements are intended primarily for population-level studies; additional visual inspection and cluster-specific vetting may be necessary for detailed analyses of individual clusters.

\begin{table*}[]
    \centering
    \caption{Column description for the full periodic signal catalog for the open cluster and moving gorup sample. 
    This table is published in its entirety in a machine-readable format in the online journal.}
    \begin{tabular}{c|c}
         Column Name & Description  \\
         \hline\hline
         Name & Cluster Name from \cite{Hunt2023} \\
         logAge16 & log(age/yr) 16$^{\rm th}$-percentile from \cite{Hunt2023} \\
         logAge50 & log(age/yr) 50$^{\rm th}$-percentile from \cite{Hunt2023} \\
         logAge84 & log(age/yr) 84$^{\rm th}$-percentile from \cite{Hunt2023} \\
         Prot\_ztf & Measured raw ZTF \prot\ \\
         Prot\_asassn & Measured raw ASAS-SN \prot\ \\
         survey & Where the \prot\ is measured from \\
         Prot\_TESS & TESS \prot\ from \cite{Boyle2026} and \cite{Kounkel2022} \\
         Prot\_lit & literature \prot\ from \cite{vanlane2025} \\
         Prot\_ground & \prot\ after correcting for failure modes (recommend using) \\ 
         m\_ztf\_correct\_tess & Whether the TESS \prot\ is used to correct for the ZTF failure modes \\
         m\_ztf\_correct\_vanlane25 & Whether the \cite{vanlane2025} \prot\ is used to correct for the ZTF failure modes \\
         m\_ztf\_correct & Whether a ZTF failure mode is corrected \\
         m\_asassn\_correct\_tess & Whether the TESS \prot\ is used to correct for the ASAS-SN failure modes \\
         m\_asassn\_correct\_vanlane25 & Whether the \cite{vanlane2025} \prot\ is used to correct for the ASAS-SN failure modes \\
         m\_asassn\_correct & Whether a ASAS-SN failure mode is corrected \\
         Ro & rossby number calculated using convective turnover time formula from \cite{See2025}\\
         GaiaDR3 & Gaia DR3 source ID \\
         ra & Gaia DR3 right ascension \\
         dec & Gaia DR3 declination \\
         pmRA & Gaia DR3 proper motion in the ra direction \\
         pmDE & Gaia DR3 proper motion in the dec direction \\
         phot\_g\_mean\_mag & Gaia DR3 $G$ magnitude \\
         phot\_rp\_mean\_mag & Gaia DR3 $G_{\rm RP}$ magnitude \\
         phot\_bp\_mean\_mag & Gaia DR3 $G_{\rm BP}$ magnitude \\
         parallax & Gaia DR3 parallax \\
         ruwe & Gaia DR3 re-normalized unit weight error \\
         binary & Gaia DR3 binary identification \\
         Var\_info & eclipsing binary (\texttt{EB}) or instability strip pulsator  (\texttt{Inst})\\
         Var\_source & literature source for \texttt{Var\_info}\\
         FAP & false alarm probability reported from \texttt{astropy.LombScargle}\\
         GOLD & whether the star is in the GOLD sample (\autoref{subsec:gold})\\
         Inspection & visual inspection results (\autoref{subsec:period_measure})

    \end{tabular}
    \label{tab:1}
\end{table*}

\subsection{The Gold sample}\label{subsec:gold}
Based on the visual inspection results mentioned in \autoref{subsec:period_measure}, I created the GOLD sample where periods are most likely reliable and the membership probabilities are high.
The GOLD sample is constructed by selecting stars with $\log$(FAP) $< -5$, with the number of epoch $>$ 200, and the membership probability reported by \cite{Hunt2023} $>$ 0.7.
This sample include $\sim$10,000 stars.

\begin{figure}
    \centering
    \includegraphics[width=\columnwidth]{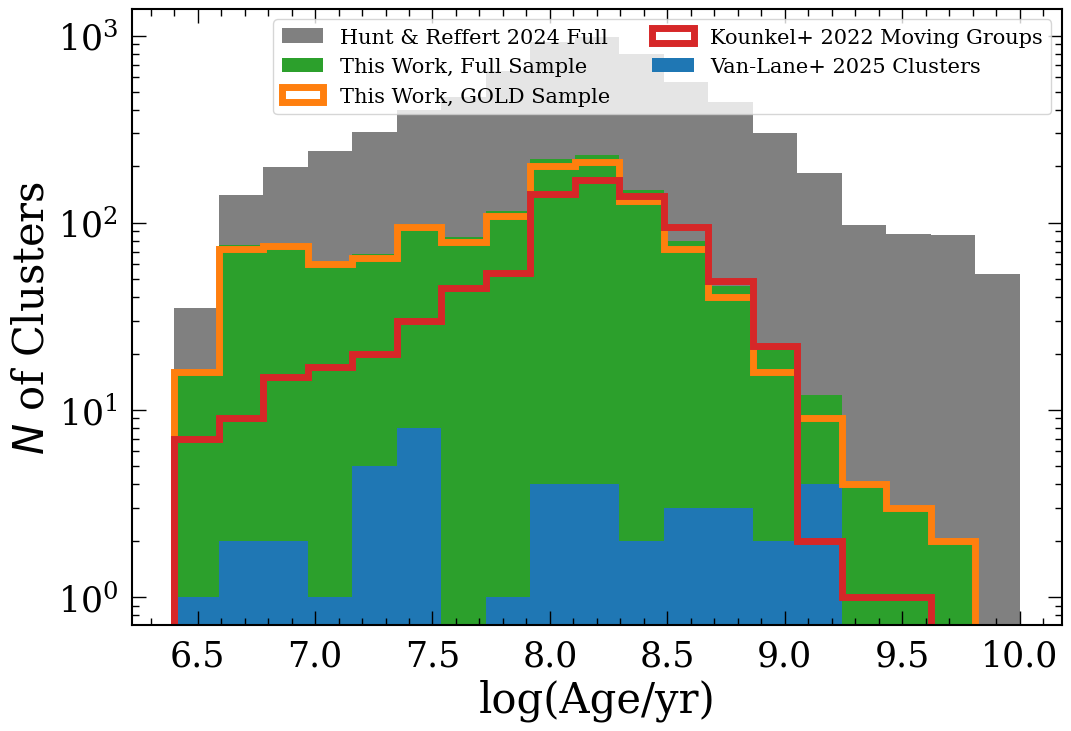}
    \caption{Age distribution of the clusters and moving groups for the full catalog \citep[gray;][]{Hunt2023}, those with rotation measurements from this work (green), the GOLD sample as described in \autoref{subsec:gold} (orange), and those with previously available rotation-period measurements from TESS (blue and red), compiled from \cite{vanlane2025} and \cite{Kounkel2022}. 
    The new sample provides a more uniform coverage in age, increasing the number of clusters with age $<$100 Myr and $>$ 1 Gyr. 
    Some of these age regimes are particularly important for understanding rotational evolution where stars younger than $\sim$10 Myr can retain significant memory of their initial rotation conditions and stars at $\sim$20--100 Myr are expected to undergo substantial core--envelope decoupling for G dwarfs. }
    \label{fig:5}
\end{figure}

\autoref{fig:5} shows the age distribution of the full cluster sample, the full rotation sample, the GOLD sample, and clusters with previously known rotation periods \citep{vanlane2025}. 
The GOLD sample includes 1,249 clusters, compared to 1,365 clusters in the full rotation sample and 43 clusters in \cite{vanlane2025}. 
This new rotation-period catalog increases the number of clusters with measured rotation periods by more than a factor of 10 and provides substantially more uniform age coverage, filling important gaps for clusters younger than 10 Myr and between 20--100 Myr, when stars can retain memory of their initial rotation conditions and core--envelope decoupling remains important for solar-type stars.

\begin{figure*}
    \centering
    \includegraphics[width=\textwidth]{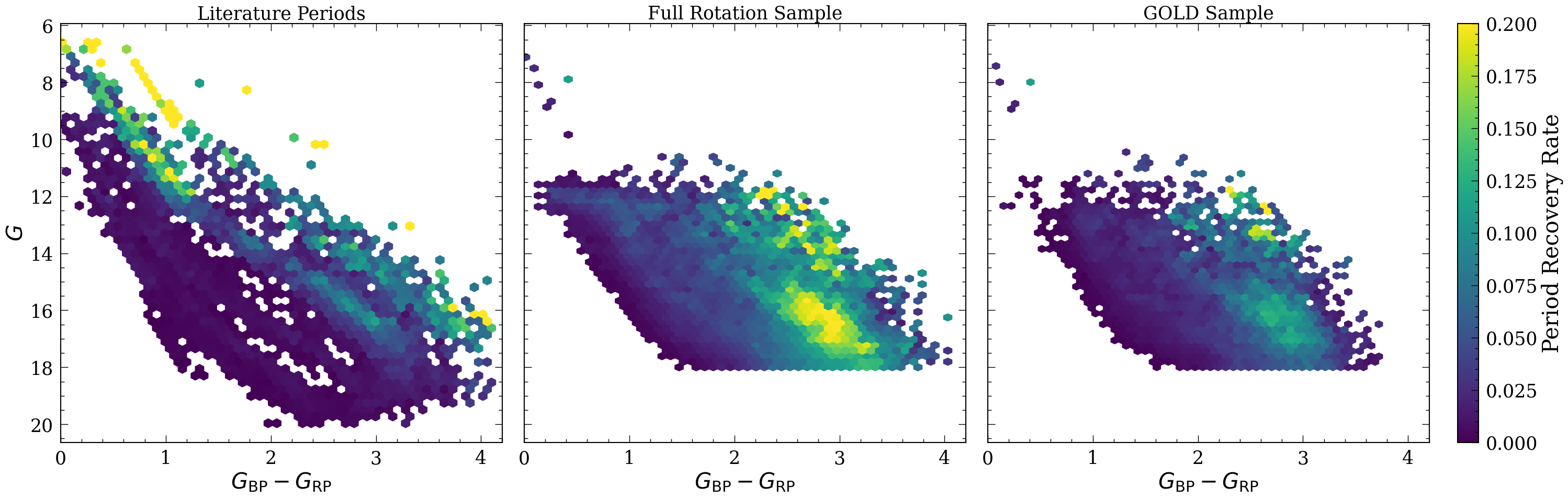}
    \caption{The period recovery fraction for the literature catalogs (left), the full rotation catalog from this work (middle), and the GOLD sample (right) in apparent Gaia $G$ magnitude--\bprp\ space. 
    The cutoffs at $G\sim18$ and 12 mag indicate the faint and saturation limits for ZTF, respectively.
    The recovery fraction is highest for K/M dwarfs, as expected given their typically higher photometric variability associated with their faster rotation. 
    The strips in the left panel reflect the small number of clusters with reported periods in the literature, whereas the full sample presented in this work provides more continuous coverage across $G$--\bprp\ space as a substantially larger number of clusters are included.
    In addition, the recovery fraction in \cite{Kounkel2022} peaks at \bprp$\sim$1 rather than among K/M dwarfs, highlighting the different sensitivities of the two instruments.}
    \label{fig:5.1}
\end{figure*}

Finally, \autoref{fig:5.1} shows the recovery fraction in Gaia $G$ versus \bprp\ space for literature-reported periods (left), the full rotation-period catalog from this work (middle), and the GOLD sample (right). 
Since the sample is dominated by signals detected from ZTF, the hard cutoff at $G=18$ mag reflects the ZTF magnitude limit imposed in \autoref{subsec:period_measure}, while the cutoff at the bright end reflects the ZTF saturation limit. 
The small number of clusters in the literature rotation catalog is visible as distinct strips due to their different distance moduli, whereas my catalog shows a more continuous distribution across color--magnitude space because it includes a substantially larger number of clusters.
The recovery fraction ranges from 0--20\%, with higher recovery rates in the K/M-dwarf regime, consistent with their shorter rotation periods and, consequently, higher photometric amplitudes.
In \cite{Kounkel2022}, the recovery fraction peaks at \bprp$\sim$1, reflecting the different sensitivities of the two instruments.

\subsection{Variable Star Catalogs}\label{subsec:EB}
I also cross-matched my catalog with available variable star and eclipsing binary catalog from ZTF \citep{Chen2020}, ASAS-SN \citep{Christy2023}, and Gaia DR3 \citep{Rimoldini2023, Mowlavi2023}.
For this paper, I combined all information into two classes, either eclipsing binaries (EB) or classical pulsator (Inst). 
I am able to identify 2,392 (12\%) EBs and 155 (0.8\%) pulsator.
These information are included in the \texttt{Var\_info} and \texttt{Var\_source} columns in \autoref{tab:1}.

\section{Results} \label{sec:results}
\subsection{The Color--Magnitude Diagram}\label{subsec:CMD}
\begin{figure*}
    \centering
    \includegraphics[width=\linewidth]{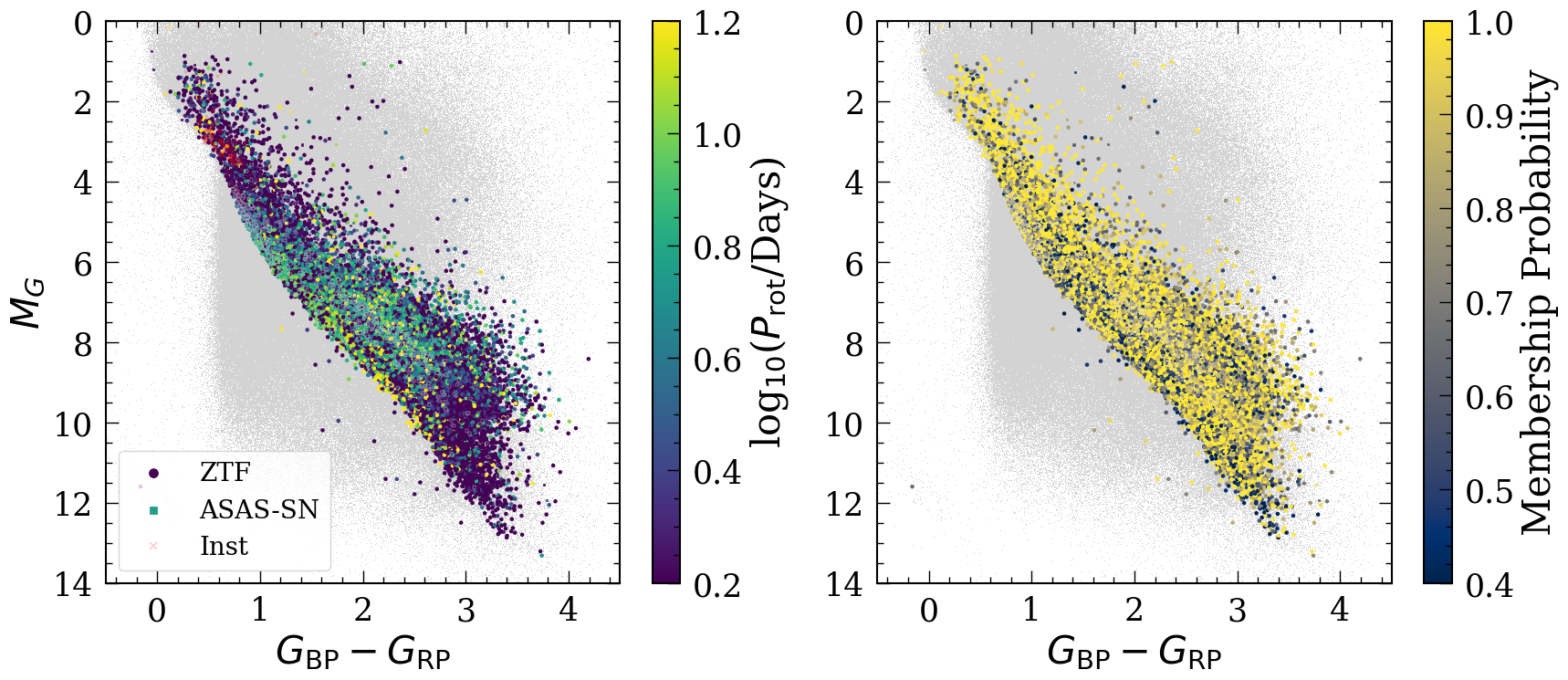}
    \caption{Color--magnitude diagram of cluster members with periodic signals detected in this work. 
    The left panel shows the measured rotation period, $\log_{10}(P_{\rm rot}/{\rm day})$, while the right panel shows the membership probability from \citet{Hunt2023}. 
    ZTF measurements are shown as colored circles and ASAS-SN measurements as gray outlined squares. 
    The EBs are excluded from this plot and the red points indicate the variable stars identified in the literature (see \autoref{subsec:EB}).
    The full cluster membership sample is shown as light-gray points in the background.
    The Gaia absolute $G$-band magnitudes are calculated using the cluster distances reported by \citet{Hunt2023}.}
    \label{fig:2}
\end{figure*}

\autoref{fig:2} shows the CMD, with the detected rotation periods shown on the left and membership probabilities on the right.
I calculate Gaia absolute $G$-band magnitudes from the apparent magnitudes using the cluster distances reported by \citet{Hunt2023}. 
All EBs are excluded from this plot and the red points indicate the variable stars identified in \autoref{subsec:EB}.
ZTF measurements are shown as colored circles, while ASAS-SN measurements are shown as gray outlined squares; the full cluster membership sample is shown as light-gray points in the background. 
As expected, stars with detected periodic variability are concentrated primarily along the pre-main sequence and main sequence.
I also identify rotation periods for several potential blue stragglers, subgiants, and giants, providing opportunities for further investigation.

Along the main sequence, the fastest rotators are concentrated at the two extremes of the color distribution, with $G_{\rm BP}-G_{\rm RP}\lesssim0.5$ and $\gtrsim2.6$. 
The bluer population lies above the Kraft break \citep{Kraft1967}, where the shallow or absent surface convection zones result in inefficient magnetic braking, while the redder population corresponds to rapidly rotating fully convective stars, consistent with previous studies. 
In contrast, the slowest rotators are concentrated toward the blue edge of the lower main sequence, where weaker magnetic activity may contribute to their distinct rotational behavior. 
Stars above the main sequence exhibit both rapid and slow rotation, potentially reflecting a mixture of pre-main-sequence stars and tidally synchronized binary systems.

\subsection{The Period Distributions}
\begin{figure*}
    \centering
    \includegraphics[width=\linewidth]{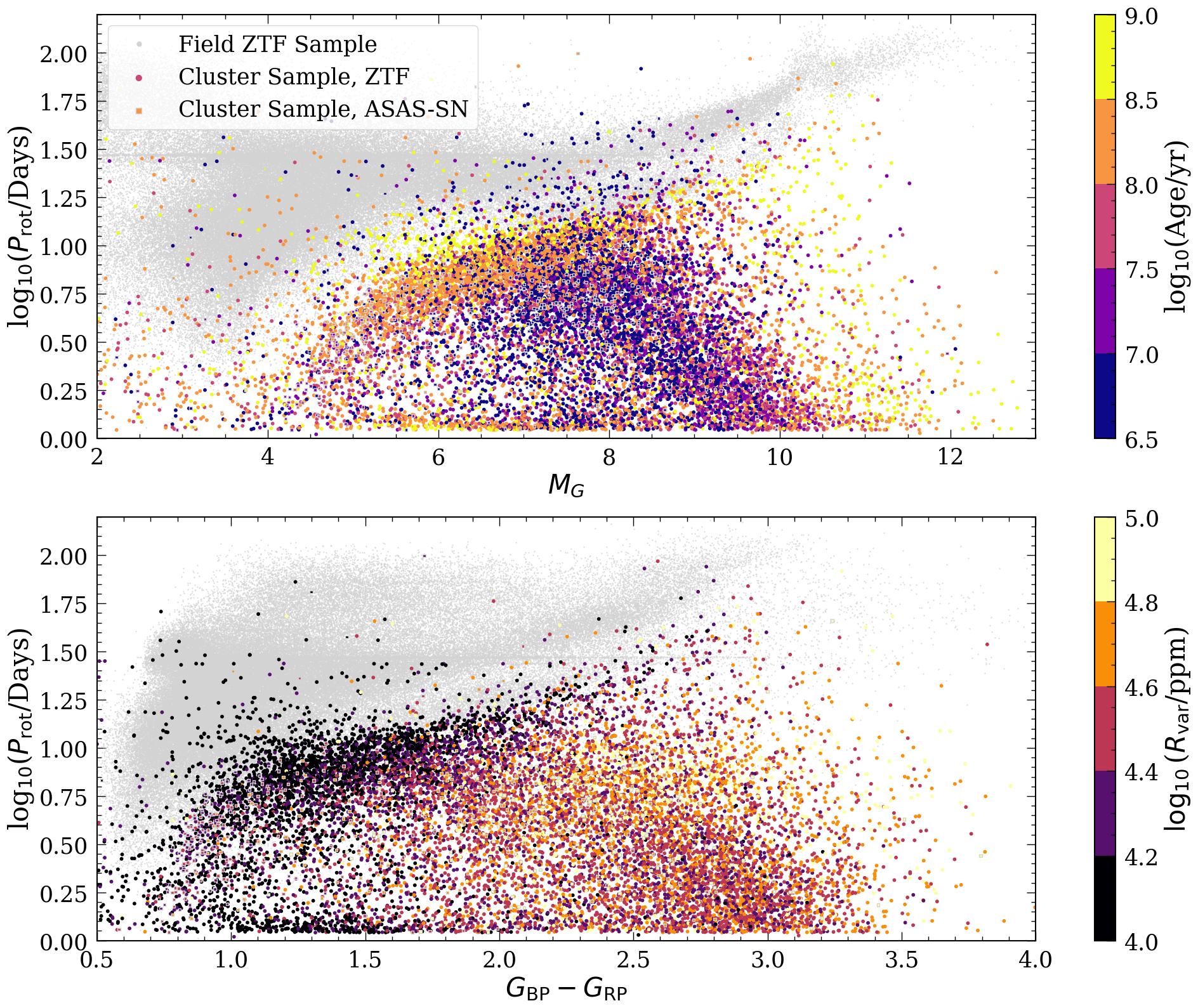}
    \caption{Main result of this work. 
    The top panel shows \logprot\ as a function of stellar absolute $G$-band magnitude, while the bottom panel shows the same rotation periods as a function of $G_{\rm BP}-G_{\rm RP}$. 
    The color scale indicates $\log_{10}(\mathrm{Age})$ on the top and photometric variability on the bottom. 
    The evolution from slow rotation during pre-main-sequence contraction to spin-down on the main sequence is visible across the diagram. 
    The EBs and variable stars are excluded from this plot, but most are distributed relatively uniformly across the hotter end of the temperature range.
    The fastest rotators with $\log_{10}(P_{\rm rot}/{\rm day})<0.25$ are likely dominated by tidally synchronized binaries. 
    The bottom panel illustrates the decrease in variability toward longer rotation periods, also as expected.}
    \label{fig:3}
\end{figure*}

\autoref{fig:3} shows the main result of our work, where \logprot\ is plotted against $M_G$ and \bprp.
Again, circles represent the ZTF stars and the squares with light gary outlines represent the ASAS-SN sample.
The points are colored by age in the top plot and the photometric variability (\rvar).
The photometric variability of each star is converted from magnitudes to relative flux units using$R_{\rm var} = 9.21 \times 10^5 \times\sqrt{\mathrm{std}(\tilde{m})^2-\mathrm{RMS}(m_{\rm err})^2}\ {\rm ppm}$, where $\tilde{m}$ is the normalized magnitude light curve, $m_{\rm err}$ denotes the corresponding photometric uncertainties, and RMS is the root-mean square.
The background light gray points show the field sample, taken literature catalogs with periods measured from ZTF \citep{Lu2022, Lu2024}, Kepler \citep{McQuillan2014, Santos2019} and TESS  \citep{Boyle2026}.
The EBs and variable stars are excluded from this plot, but most are distributed relatively uniformly across the hotter end of the temperature range.

The top panel shows the expected evolution of stellar rotation with age where stars initially contract toward the main sequence and spin up, followed by gradual spin-down as they lose angular momentum. 
This behavior is most apparent for stars with $M_G$ between 6 and 10, where the slowest rotators are the youngest stars that have not yet completed their contraction onto the main sequence.
The ``waterfall'' feature previously identified in cluster rotation catalogs is also visible among the lowest-mass stars \citep[e.g.,][]{Curtis2020, Douglas2016}. 
The fastest rotators, with $\log_{10}(P_{\rm rot}/{\rm day})<0.25$, are likely dominated by tidally synchronized binaries.
The bottom panel shows the same rotational evolution as a function of $G_{\rm BP}-G_{\rm RP}$, along with the expected decrease in photometric variability toward longer rotation periods or hotter stars, consistent with the connection between stellar rotation and magnetic activity.
Compared to the TESS moving-group catalog presented in \cite{Kounkel2022}, this catalog extends the period range beyond 12 days, providing a complementary catalog of stars with longer rotation periods.

\subsection{The Rotation-Activity Relations}\label{subsec:RoRvar}
\begin{figure*}
    \centering
    \includegraphics[width=\linewidth]{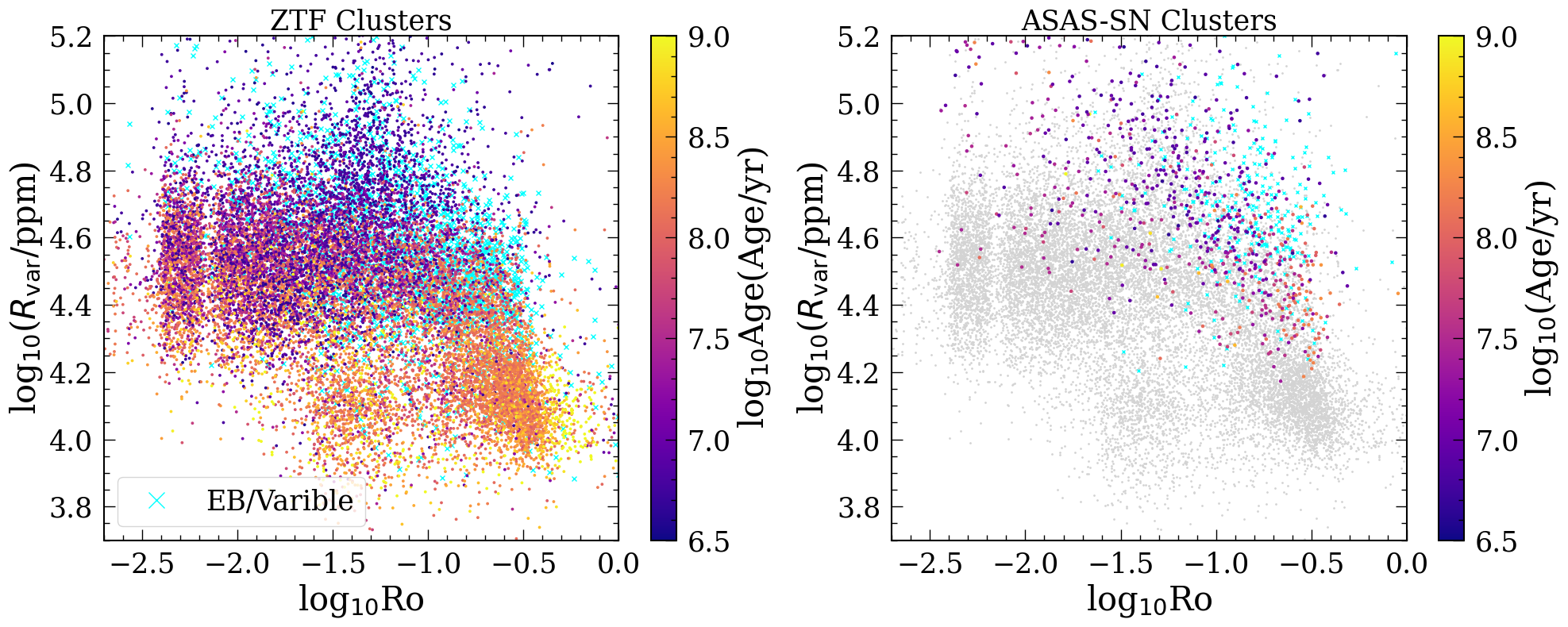}
    \caption{Rotation--activity relation for cluster members in ZTF (left) and ASAS-SN (right). 
    Photometric variability, $R_{\rm var}$, is shown as a function of Rossby number, $Ro=P_{\rm rot}/\tau_c$, with points colored by cluster age. 
    Literature-identified EBs and variable stars are shown as cyan points and generally lie above the normal activity trend.
    The light-gray points in the background in the right panel show the full ZTF cluster sample for comparison. 
    The gap near $\log_{10}Ro\sim-2.25$ results from the removal of periods associated with the 1-day systematic. 
    The approximately flat relation at Ro$<0.1$ corresponds to the saturated activity regime, while the decline in \rvar\ toward larger Rossby numbers traces the unsaturated regime. 
    The youngest stars with enhanced variability near $\log_{10}$R$\sim-1.25$ and the relatively old, rapidly rotating stars with low variability around the same Ro are highlighted as distinct populations warranting further investigation.}
    \label{fig:4}
\end{figure*}

\autoref{fig:4} shows the rotation--activity relation for ZTF (left) and ASAS-SN (right) as a function of Rossby number, $\mathrm{Ro} = P_{\rm rot}/\tau_c$, where $P_{\rm rot}$ is the rotation period and $\tau_c$ is the convective turnover time. 
I estimate $\tau_c$ from the stellar effective temperature using the empirical relation presented by \citet{See2025}. 
The Rossby number provides a physically motivated measure of the coupling between stellar rotation and convection and is therefore more directly related to magnetic activity than rotation period alone.
In particular, stellar magnetic fields are thought to be generated by dynamo processes operating within the convection zone, making the ratio of the rotation timescale to the convective timescale an important parameter for characterizing magnetic activity \citep[e.g.,][]{Noyes1984}. 
I estimate effective temperatures from \bprp\ using the empirical color--temperature relation of \citet{Curtis2020}.

The ZTF cluster stars are also shown as light-gray points in the background of the right panel for comparison. 
All detected periodic variables are colored according to their cluster ages. 
The EBs and variable stars identified in the literature are shown as cyan points, with most exhibiting elevated activity relative to the normal trend.
The gap at $\log_{10}Ro\sim-2.25$ results from the exclusion of periods close to the 1-day systematic. 
Overall, the rotation--activity relation exhibits the expected behavior where stars with Ro$<0.1$ show an approximately flat relation between Rossby number and photometric variability, corresponding to the saturated activity regime, while stars with Ro$>0.1$ exhibit a power-law decline in variability toward larger Rossby numbers, characteristic of the unsaturated regime. 
This behavior is consistent with the classical rotation--activity relation observed in previous studies \citep[e.g.,][]{Wright2011, See2025}.

Several additional features are noteworthy. 
The youngest, bluest stars appear to reach particularly high levels of photometric variability near $\log_{10}$Ro$\sim-1.25$, with $\log_{10}(R_{\rm var})\gtrsim4.8$. 
Examination of their Gaia stellar parameters reveals no obvious differences between these stars and the broader cluster population beyond their younger ages and lower temperatures.
Most of these objects lie near the fully convective boundary, with temperatures of $\sim3500$ K, which also coincides with the maximum in the convective turnover time adopted in the empirical relation.
One possibility is that the empirical relation used to estimate the convective turnover time is inaccurate in this temperature regime. 
However, even shifting their inferred Rossby numbers would not remove their elevated activity relative to the rest of the sample, suggesting that this feature is not solely an artifact of the adopted turnover times. 
Alternatively, the enhanced activity may reflect changes in the stellar dynamo associated with the transition from partially to fully convective interiors at very young ages. 

Another notable feature appears at the opposite end of the activity distribution at a similar Rossby number. 
These stars are relatively old but exhibit low photometric variability, with $\log_{10}(R_{\rm var})\lesssim4.2$. 
Further examination shows that they are essentially counterparts to the young, highly variable population discussed above --- they are higher-mass stars that retain surface convection zones but rotate unexpectedly rapidly for their old ages. 
Their rapid rotation could potentially be maintained through tidal synchronization in binary systems, placing them in a similar region of parameter space as the short-period binary candidates discussed above. 
However, their renormalized unit weight error (RUWE) distribution is not significantly different from that of the broader sample,  providing no clear astrometric evidence for an elevated binary fraction. 
In addition, they exhibit lower photometric variability than expected from the nominal trend, in contrast to the EBs.
Additional observations and independent measurements of stellar temperatures, convective turnover times, and magnetic activity will be needed to determine the formation pathway of these outliers in the period-activity relation.

\section{Potential Applications and Future Directions}\label{sec:Uses}
Here I highlight several potential research directions enabled by this large cluster rotation catalog:

\begin{itemize}
\item Metallicity dependence of stellar spin-down.
Although most open clusters have near-solar metallicities, the sample provided in this work spans a substantial range in both age and metallicity, with most clusters having $-0.4 \lesssim {\rm [Fe/H]} \lesssim 0.2$ according to Gaia XP metallicity \citep{Andrae2023}.
This provides a unique opportunity to empirically investigate the efficiency of angular-momentum loss as a function of chemical compositions in open clusters without having to take into account age affects in the field sample and to test the metallicity dependence of current gyrochronology and stellar-wind models.

\item Recalibrating gyrochronology and angular-momentum loss.
Unlike existing space-based rotation catalogs, the sample presented in this work is not limited by the short observing baselines of individual K2 campaigns or TESS sectors. 
The multi-year baselines of ZTF and ASAS-SN enable measurements of longer rotation periods, providing an opportunity to extend and recalibrate gyrochronology relations into a regime that is poorly constrained by existing cluster samples.

\item Connecting stellar rotation, age, and activity.
The combination of rotation periods, cluster ages, and photometric variability provides a large empirical dataset for studying the coupled evolution of stellar rotation and magnetic activity.
In particular, this sample can be used to investigate the evolution of the rotation--age--activity relation and to test how the convective turnover time and magnetic activity depend on stellar mass and age.

\item Investigating rotation--activity outliers.
Although the overall rotation--activity relation follows the expected saturated and unsaturated regimes, several populations deviate significantly from the general trend, as discussed in \autoref{subsec:RoRvar} and shown in \autoref{fig:4}. 
These outliers may provide new insights into stellar dynamos, magnetic-field generation, and the transition between different interior convection regimes.

\item Understanding slow-rotating fully convective M dwarfs. 
In Lu et al. (in prep.), we identified slow-rotating ($>35$ days) fully convective M dwarfs in the Hyades and Praesepe that cannot be fully explained by field contamination. 
However, it remains unclear when these slow rotators emerge. If slow rotators are already present in younger clusters, their long periods may reflect their initial rotation periods. 
If they emerge only at later ages, their presence would instead point to stronger magnetic braking.

\end{itemize}

\section{Conclusion} 
I present a catalog of $\sim20,000$ periodic photometric signals for members of nearby \NClusters\ open clusters and moving groups within 1.5 kpc, measured using multi-year light curves from ZTF and ASAS-SN.
This catalog contains \NZTFvet\ signals from ZTF and \NASASSNvet\ from ASAS-SN, more than doubling the number of previously available rotation-period measurements for these populations. 
By combining the long temporal baselines of ground-based time-domain surveys with existing space-based measurements, I extend rotation-period measurements to stars and timescales that are difficult to probe with individual K2 campaigns and TESS sectors.

I develop a systematic procedure to identify and correct periodogram failure modes arising from uneven ground-based temporal sampling. 
After correction, 97\% of the ZTF and 98\% of the ASAS-SN periods agree with literature measurements or their half- or double-period harmonics, demonstrating the reliability of the period measurements at the population level. 
I additionally identify and manually vet stars in old clusters ($>1$ Gyr) and statistical outliers in the rotation--color--magnitude--activity parameter space, further improving the reliability of the catalog. 
Because the catalog is designed primarily for population-level studies, additional cluster-specific vetting may be appropriate for detailed investigations of individual systems.
Using guidance from the manually vetted sample, I constructed the GOLD sample of $\sim10,000$ sources, for which the measured periods are most reliable and cluster membership probabilities are high. 
We cross-matched this catalog with existing EB and variable-star catalogs from the literature and identified $\sim2,300$ (12\%) EBs and 155 (0.8\%) classical pulsators.

The large and heterogeneous sample reveals several expected features of stellar rotation. 
Rotation periods increase and photometric variability decreases systematically with cluster age as stars evolve from pre-main-sequence contraction and spin-up to main-sequence angular-momentum loss and spin-down (\autoref{fig:3}).
The catalog also reproduces the classical rotation--activity relation where stars with $Ro\lesssim0.1$ occupy the saturated regime, while photometric variability decreases toward larger Rossby numbers in the unsaturated regime (\autoref{fig:4}). 
Several populations depart from this general relation, including young, highly active stars near the fully convective boundary and old stars that rotate unexpectedly rapidly. 
These populations provide interesting targets for future investigations of stellar dynamos, magnetic activity, and binary evolution.

In addition, this catalog includes clusters spanning ages and metallicities that are poorly represented in previous catalogs, particularly non-solar-metallicity clusters and clusters younger than 10 Myr or between 20--100 Myr, where stars retain memory of their initial conditions and G dwarfs experience core--envelope decoupling, respectively.
The long time baselines and broad sky coverage of ZTF and ASAS-SN are particularly valuable for constructing a more complete cluster sample, providing a new empirical foundation for studying stellar angular-momentum evolution across a broad range of ages, masses, and metallicities.

\begin{acknowledgments}
I thank Marc Pinsonneault for the helpful discussion.
This work has made use of data from the European Space Agency (ESA) mission Gaia,\footnote{\url{https://www.cosmos.esa.int/gaia}} processed by the Gaia Data Processing and Analysis Consortium (DPAC).\footnote{\url{https://www.cosmos.esa.int/web/gaia/dpac/consortium}} 
Funding for the DPAC has been provided by national institutions, in particular the institutions participating in the Gaia Multilateral Agreement.
This research also made use of public auxiliary data provided by ESA/Gaia/DPAC/CU5 and prepared by Carine Babusiaux. 
This research has also made use of NASA's Astrophysics Data System.
\end{acknowledgments}

%

\vspace{5mm}
\facilities{ASAS-SN \citep{asassn2014}, PO:1.2m \citep[ZTF][]{ztfdata, ztftime}, Gaia \citep{gaia, Gaiadr3}, TESS \citep{tess}}

\software{astropy \citep{astropy:2013, astropy:2018, astropy2022},   \texttt{lightkurve} \citep{lightkurve}, \texttt{Matplotlib} \citep{matplotlib}, \texttt{NumPy} \citep{Numpy}, \texttt{Pandas} \citep{pandas}, \texttt{ChatGPT} \citep{openai2025chatgpt}, ASAS-SN Sky Patrol \citep{Hart2023}.}



\bibliography{sample631}{}
\bibliographystyle{aasjournal}



\end{document}